\documentclass[final]{PoS}
\usepackage{shrthnds}
\usepackage{subfigure}
\usepackage{setspace}
\usepackage{multirow}

\newcommand{\gpZ}{g g \to g \gamma Z}

\newcommand{\ppg}{g g  \to \gamma \gamma g}

\title{Multi Vector Boson Production via Gluon Fusion at the LHC}

\ShortTitle{Multi Vector Boson Production via Gluon Fusion at the LHC}

\author{Pankaj Agrawal and \speaker{Ambresh Shivaji} \\ 
       Institute of Physics, Sachivalaya Marg, Bhubaneswar 751 005, India\\
       E-mail: \email{agrawal@iopb.res.in}, \email{ambresh@iopb.res.in}}

\abstract{At the LHC, the processes with several particles in the final
   state occur routinely. These processes provide a new domain to test the Standard
   Model of particle physics. Due to large gluon luminosity at the LHC,
   gluon initiated processes will be quite important. We study
   the production of multi-vector bosons via gluon fusion. In particular
   we consider the process $ g g \to g \gamma Z $ and compute its cross-section at
   hadron colliders. }

\FullConference{10th International Symposium on Radiative Corrections (Applications of
                Quantum Field Theory to Phenomenology)\\
                 September 26-30, 2011\\
                Mamallapuram, India}

\begin{document}
\section{Introduction}

   The Large Hadron Collider (LHC) has begun its operation. The Standard Model (SM)
  seems to be in excellent agreement with the collected data. There has been
  searches for the hints of physics beyond the SM such as
  supersymmetry, large extra dimensions, etc. But, as of now, there is no clear
  evidence \cite{cern}. The search for the Higgs boson is also continuing. With the
  lack of signals for beyond the SM scenarios, there is a
  need to look for the SM processes that were not accessible
  at the Tevatron. Most of such processes have several particles in the
  final state, and/or occur at the higher order. Such processes can also
  contribute to the background to the new physics signals. One such class
  of processes is multi vector boson production in association with
  one or more jets.\\

\begin{figure}[h!]
\bc
\includegraphics[width=0.7\textwidth]{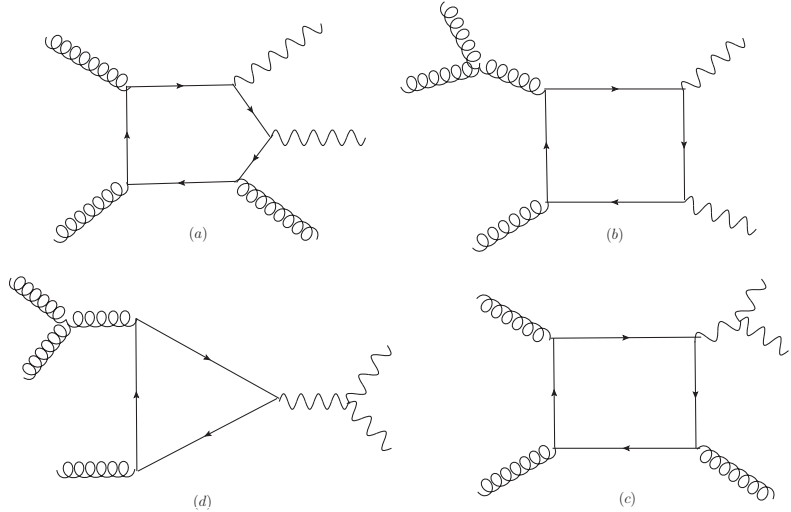}\label{fig:VVg}
\ec
\caption{The prototype diagrams for the processes $gg \to V V^\prime g$. 
The wavy lines represent the appropriate combination of
the $\gamma, W, \rm{or}\; Z$ boson. The last two classes $(c)$ and $(d)$ are relevant to $WWg$
production only.}
\end{figure}

    At the LHC centre of mass energy, the collider has another useful feature.
   In the proton-proton collisions, the gluon luminosity can be quite significant.
   It can even dominate over the quark luminosity in certain kinematic domains.
   Therefore at the LHC, loop mediated gluon fusion processes can be important.
   Di-vector boson production via gluon fusion have been studied by many authors 
   \cite{vanderBij:1988fb,Glover:1988rg,Glover:1988fe,Campbell:2011bn,Campbell:2011cu}.
   We consider a class of processes $gg \to V V^\prime g$, where $V$ and $V^\prime$
   can be any allowed combination of electroweak vector bosons. At the leading order, these processes
   receive contribution from quark loop diagrams. The prototype diagrams are displayed
   in Fig (\ref{fig:VVg}). The computation for the process $\ppg$ has already been performed 
   \cite{deFlorian:1999tp,Agrawal:1998ch}.
   In this contribution to the proceedings,
   we particularly focus on a specific process $\gpZ$. We report our results of the cross-section calculation
   for this process at hadron colliders. After this brief introduction,
   in the next section, we describe our process and discuss the structure of the amplitude. 
   In the section 3, we provide details on our calculation and numerical checks. In the section 4, we present
   brief numerical results. We conclude in the last section.

\section{The Process}

    The process $\gpZ$ contributes to the cross-section of the
   process $pp \to \gamma Z + \rm{jet}$. This process can be
   a background to new physics scenarios, in particular to
   the technicolor scenarios. Although the process $\gpZ$ occurs
   at the one loop level, its cross-section can be significant
   due to large gluon luminosity at the LHC. This one-loop
   process receives contribution from two main classes of quark loop diagrams
   -- pentagon and box type, as shown in $(a)\;\rm{and}\;(b)$ of Fig (\ref{fig:VVg}). 
   Box class of diagrams are due to the triple gluon vertices
   and can be further divided into three sub-classes. This sub-classification has it's 
   own physical importance as we will see later. 
   Other diagrams can be obtained by suitable permutation of external
   legs. For each quark flavour, there are 24 pentagon-type and $3\times6$ (= 18) box-type diagrams. 
   Due to Furry's theorem, only half of the 42
   diagrams are independent. We work with all six quark flavours. Except $t$-quark we treat 
   all other quarks massless. Our one loop process, being at leading order, is expected to be finite.\\

  The amplitude for our process has very nice structure. Let us first consider
  the prototype box diagram. Since Z-boson has both vector and axial-vector couplings 
  with quark, one can write
  this diagram as a sum of two pieces -- vector piece and axial-vector
  piece 
\begin{equation}
 {\cal A}_{q,B}^{i} = {\cal A}_{q,B}^{i,V} + {\cal A}_{q,B}^{i,A}.
\end{equation}
 Here $i$ is the diagram index. It takes value from 1 to 18, for
 each quark flavour. Furthermore, $q$ is the flavor index for the
 quark in the loop and runs from 1 to 6.\\

\begin{figure}[h!]
\bc
\includegraphics[width=0.5\textwidth]{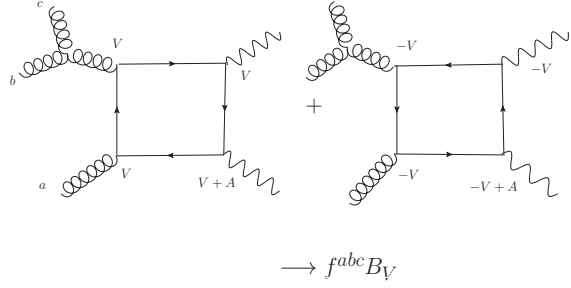}\label{fig:box}
\ec
\caption{Implication of Furry's theorem for charge-conjugated box-diagrams.}
\end{figure}

  The axial-vector piece will cancel when we sum over all box
  diagrams. This happens because we can categorize 18 box diagrams for each
  quark flavour in two parts. One part has momentum flowing in clockwise
  direction and other part has momentum flowing in the anticlockwise 
  direction. These two parts have identical vector pieces, but the axial
  vector pieces have opposite signs. It leads to the cancellation
  of axial vector pieces. Therefore, the axial-vector coupling of the Z-boson
  to the quarks does not make any contribution to the full box amplitude.
  There is pairwise cancellation between charge-conjugated fermion loop diagrams
  as shown in Fig (\ref{fig:box}). It means
\begin{equation}
 \sum_{i=1}^{18}  {\cal A}_{q,B}^{i,A} = 0.
\end{equation}

\begin{figure}[h!]
\bc
\includegraphics[width=0.7\textwidth]{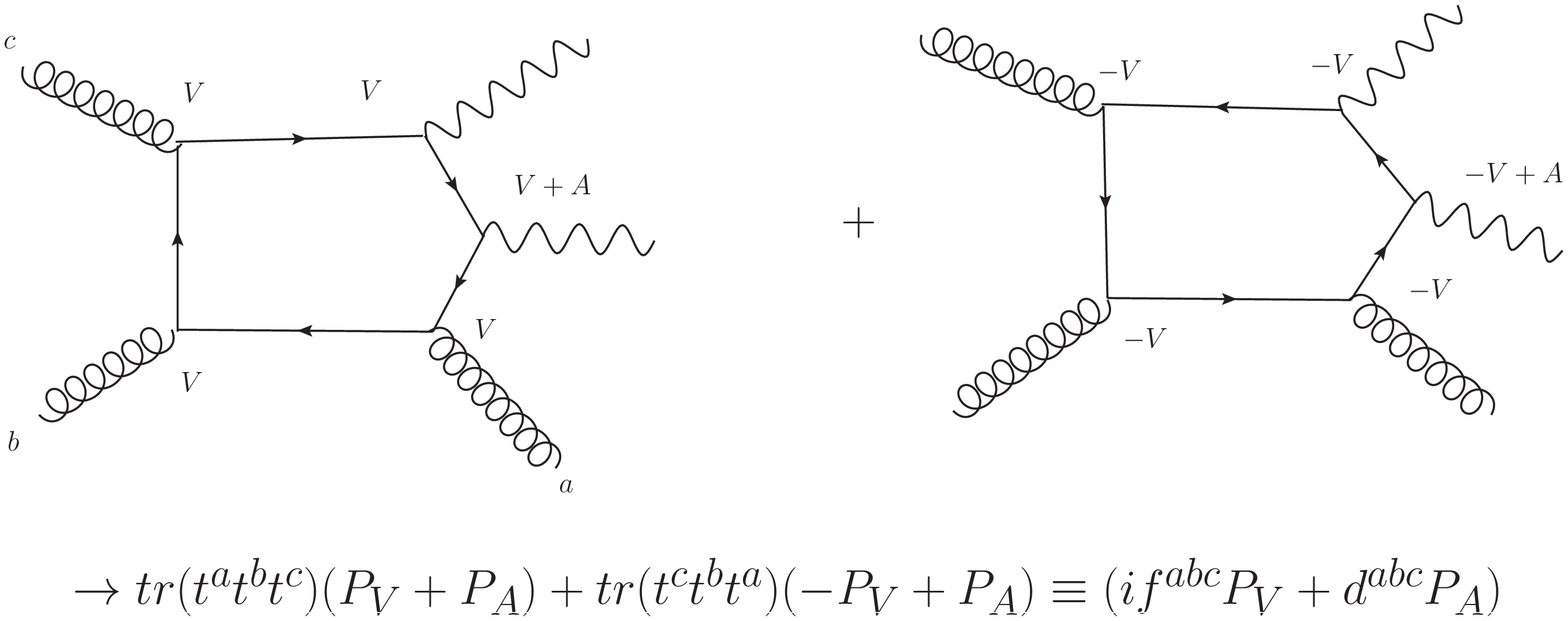}\label{fig:penta}
\ec
\caption{Implication of Furry's theorem for charge-conjugated pentagon-diagrams.}
\end{figure}

    The situation is different for the pentagon diagrams. We can again write
  the contribution of each pentagon diagram as the sum of vector and axial-vector pieces
\begin{equation}
 {\cal A}_{q,P}^{i} = {\cal A}_{q,P}^{i,V} + {\cal A}_{q,P}^{i,A}.
\end{equation}
   Here $i$ is the diagram index. It runs from 1 to 24 for each quark flavour.
   The flavour index $q$ runs from 1 to 6.
    Without color factor, the ${\cal A}_{q,P}^{i,V}$ for the diagrams with
  opposite direction of the flow of momentum have opposite signs,
  while the ${\cal A}_{q,P}^{i,A}$ have same signs. Because of the opposite
  signs, the sum of the vector pieces of all pentagrams have same color
  factor, as the sum of box diagrams. This color factor is antisymmetric
  (i.e., it is proportional to the structure constants of the SU(3)
   Lie algebra, $f_{abc}$). The color factor for the axial-vector
   piece is symmetric (i.e. proportional to $d_{abc}$). This has been shown 
in Fig (\ref{fig:penta}).
Thus the full amplitude has following general structure

\begin{eqnarray}
M(\gpZ) =  \sum_{q=1}^{6}\left[\sum_{i=1}^{18}  {\cal A}_{q,B}^{i,V} +  \sum_{i=1}^{24} \left( {\cal A}_{q,P}^{i,V} +  
            {\cal A}_{q,P}^{i,A}\right)\right]  \\
\Rightarrow M^{abc}(\gpZ) = i \frac{f^{abc}}{2} M_V + \frac{d^{abc}}{2} M_A \\
M_V = {\cal A}_P^V - {\cal A}_B^V = \frac{e^2g_s^3}{{\rm sin}{\theta_w} {\rm cos}{\theta_w}} [\left(\frac{7}{12} - \frac{11}{9} {\rm sin}^2{\theta_w}\right) M_V^{(0)}
 & \nonumber \\
                                                              + \left(\frac{1}{6} - \frac{4}{9} {\rm sin}^2{\theta_w}\right) M_V^{(t)}]\\ 
M_A = {\cal A}_P^A = (-)\frac{e^2g_s^3}{{\rm sin}{\theta_w} {\rm cos}{\theta_w}} \left[\frac{7}{12} M_A^{(0)} + \frac{1}{6} M_A^{(t)}\right].
\end{eqnarray}
$M_{V,A}$ are color ordered amplitudes for the vector and axial-vector parts of the amplitude.
${\cal A}_B^V\;\rm{and}\; {\cal A}_P^{V,A}$ include full box and pentagon contributions for all six quark flavours.
$M_{V,A}^{(0)}$ and $M_{V,A}^{(t)}$ represent light quark and $t$-quark contributions respectively. The 
structure of the amplitude suggests that the vector and axial-vector contributions should be separately gauge invariant.
Moreover due to the color structure when we square the amplitude, the interference between the axial and vector 
contribution vanishes, $i.e.$
\begin{eqnarray}
|M(\gpZ)|^2 = \left( 6 |M_V|^2 + \frac{10}{3} |M_A|^2 \right).
\end{eqnarray}
     
Therefore the cross-section of the process is an incoherent sum of the vector and axial-vector contributions. 
Hereafter we will consider only the vector contribution, $M_V$.

\section{Calculation and Numerical Checks}

 For each class of diagrams, we write down the prototype amplitudes
using SM Feynman rules. The amplitude of all other 
  diagrams are generated by appropriately
  permuting the external momenta and polarizations in our code. The quark loop traces are calculated 
 in FORM in $n$-dimensions \cite{Vermaseren:2000nd}. The amplitude contains tensor loop integrals.
In the case of pentagon-type diagrams, the most complicated integral
is rank-5 tensor integral ($E^{\mu \nu \rho \sigma \delta}$); while
for the box-type diagrams, rank-4 tensor integral ($D^{\mu \nu \rho \sigma}$) is
the most complicated one 

\begin{equation}
E^{\mu \nu \rho \sigma \delta} = \int \frac{d^n k}{(2 \pi)^n} \frac{k^{\mu} k^{\nu} k^{\rho} k^{\sigma} k^{\delta}}
 {N_0 N_1 N_2 N_3 N_4}\,,
\end{equation}

\begin{equation}
D^{\mu \nu \rho \sigma} = \int \frac{d^n k}{(2 \pi)^n} \frac{k^{\mu} k^{\nu} k^{\rho} k^{\sigma}}
 {N_0 N_1 N_2 N_3}\,.
\end{equation}
Here, $N_i = k_i^2 - m_q^2 + i\ep$ and $k_i$ is the momentum of the $i^{th}$ internal line in the
corresponding scalar integrals.
$n=(4-2\ep)$ and $m_q$ is the mass of the quark in the loop.
Five point tensor and scalar integrals are written in terms of box tensor and 
scalar integrals using 4-dimensional Schouten Identity. For example, a five point scalar integral, 
in 4-dimensions can be expressed in terms of five box scalar integrals \cite{vanNeerven:1983vr}

\begin{equation}
 E_0(0,1,2,3,4) = \sum\limits_{i=0}^4 c_i D_0^{(i)},
\end{equation}
where $D_0^{(i)}$ is the box scalar integral obtained after removing the $i^{th}$ propagator
in $E_0$. It can be argued that any $O(\ep)$ correction to the above relation in $n=(4-2\ep)$-
dimensions does not contribute to our amplitude.  
The box tensor integrals are reduced into the standard scalar integrals -- $A_0$, $B_0$, $C_0$ and $D_0$
 using fortran routines \cite{Agrawal:1998ch} that follows from the reduction scheme
developed by Oldenborgh and Vermaseren \cite{vanOldenborgh:1989wn}.
 Thus we require box scalar integrals with two massive external legs, at the most.
The scalar integrals with massive internal lines are calculated using OneLoop library \cite{vanHameren:2010cp}.
Because of very large and complicated expression of the amplitude, we calculate the amplitude numerically before 
squaring it. This requires numerical evaluation of the polarization vectors of gauge bosons.
We choose real basis, instead of helicity basis for the polarization vectors to calculate the amplitude.
This is to reduce the size of compiled program and the time taken in running the code. \\

   The process $\gpZ$ is a leading order one-loop process. It should be both UV as well as IR finite. 
  However, individual diagrams may be UV and/or IR divergent. IR divergence is relevant to only light 
  quark cases.
All these singularities are encoded in various scalar integrals. 
 To make UV and IR finiteness check on our amplitude we have derived all 
the required scalar integrals (upto box scalar integrals with two massive external legs)
analytically following t`Hooft and Veltman \cite{'tHooft:1978xw}. We regulate the UV divergence 
of the scalar integrals using dimensional regularization and infrared singularities by using small 
quark mass (the mass regularization).  Our results 
are in agreement with those given in \cite{Duplancic:2000sk}.
 Following are the details of various checks made on our amplitude,

\begin{enumerate}
\item {\it UV Finiteness:}

The ultimate source of the UV divergence is the bubble scalar integral, $B_0$.
For both the massive and massless quark contributions, we
have verified that the amplitude is UV finite. Each pentagon diagram
is UV finite by itself as expected from naive power counting. The 
three classes of box diagrams are also separately UV finite.

\item {\it IR Finiteness:}

  The diagrams with massless internal quarks have mass singularities.
 Even with small quark mass (like bottom quark) these diagrams may have
 large logarithms which should cancel for the finiteness of the 
 amplitude. There can be ${\rm log}^2(m_q)$ and ${\rm log}(m_q)$ type of singular 
 terms. We have checked explicitly that such terms are absent from
 the amplitude. We have verified that the IR finiteness holds for each 
box and pentagon diagrams \cite{Shivaji:2010aq}.

\item {\it Gauge Invariance:}

The amplitude has gauge invariance with respect to three gluons, the photon and the Z-boson. This has
 been checked by replacing the polarization
vector of any of these gauge particles by its momentum ($\ve^\m(p_i) \rar p^\m_i$) 
which makes the amplitude vanish. As one would expect the pentagon and three classes of box
contributions are separately gauge invariant with respect to
the $\gamma$ and Z. For each gluon one of the three classes of box amplitudes is
separately gauge invariant and further cancellation takes place among
pentagon and the other two box contributions.

\item{\it Decoupling of heavy quarks: }

As a consistency check we have also verified that the vector part of the amplitude vanishes
in the large quark mass limit \cite{Appelquist:1974tg}. This feature of the amplitude is very closely related to it's
UV structure. The decoupling holds for each penta amplitude and also for each class of box 
amplitudes. 

\end{enumerate}

\section{Numerical Results}

       We have computed the amplitude numerically using the real polarization
    vector basis. However, given the number of diagrams, the number of
    polarization combinations and the length of the amplitude, we have to
    run the code in parallel environment using a PVM implementation of the VEGAS
    algorithm \cite{Veseli:1997hr,pvm}. We have used more than 20 cores to run the code in parallel environment.
    Still it takes more than a day time to get suitable cross-section. To
    obtain the distributions of the kinematic variables, one will have to run
    the code on more cores for a longer time.

       In such loop calculations, there is an issue of numerical instabilities
    due to large cancellations in the evaluation of the scalar and tensor integrals.
    To minimize this problem, we have used the OneLoop implementation of the
    scalar integrals, $B_{0}, C_{0}, {\rm and}\;  D_{0}$. Even with this 
    implementation, we have faced numerical instabilities in the evaluation
    of pentagon-tensor integrals. This is related to inaccurate evaluation of Gramm determinants 
    for certain phase-space points. We have used stiff cuts on the $p_T$ to
    reduce the impact of this problem. With this, we have about 1 in
    10,000 points that give rise to numerical instabilities. We need to 
    study this issue more closely. For the time being we judiciously throw away
    such points. This is a general practice. Since such points are few,
    given the nature of the integral evaluation, we don't expect significant
    impact of this aspect of the calculation. 
       
\begin{figure}[]
\bc
\includegraphics [angle=0,width=0.7\linewidth] {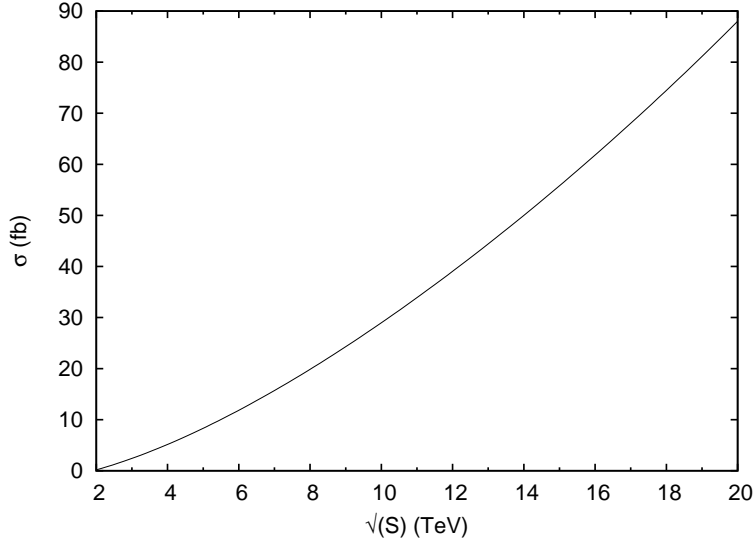}\label{fig:sigma_cme}
\ec
\caption{Variation of the cross-section ($\s$) with the centre of mass energy}
\end{figure}

In Fig (\ref{fig:sigma_cme}), we present the results of the cross-section calculation.
 These results include following kinematic cuts:
$$
P_T^{\gamma,Z,j} > 50 \; \rm{GeV},\: |\eta^{\gamma,Z,j}| < 2.5,\; R(\gamma,j) > 0.6.
$$
We have also chosen factorization and renormalization scales as
$\mu_f = \mu_R = p_T^Z$. For practical reasons, we have used CTEQ4 NLO
parton distribution functions. (This is because of the availability 
of the parametrized versions of the parton distribution functions
without the external tables.) We note that at typical LHC
energy, the cross-section would be of the order of 40 fb. Still 
one may expect few thousand of such events
after a few years of LHC operation at 14 TeV CM energy. If we relax
the cut on the $p_T$ of the photon and the jet, then the cross-section
would increase significantly. Our cut is stiff to avoid the numerical
instability. As we discussed above the axial-vector contribution will
also increase the cross-section. 

\section{Conclusions}

   We have presented the results of the cross-section calculations 
  for the process $\gpZ$. This process occurs at one-loop level
  via pentagon and box diagrams. We have made a number of checks 
    on our calculation. We have verified the cancellation of UV and mass 
    singularities. We have also checked gauge invariance with respect to
    all the gauge particles.
     We have plotted the cross-section as a function of the
    centre of mass energy. The results are for
    the vector contribution only. This contribution is gauge invariant. 
    Axial-vector
    contribution does not interfere with the vector contribution and it is
    separately gauge invariant. This axial-vector contribution can only increase the 
    cross-section. We have obtained the cross-sections
  using the parallel running of the code. With a stiff cut of about
  of 50 GeV on the transverse momentum of the photon and the jet,
  the cross-section is about 40 fb at the centre of mass energies
  of 14 TeV.

\section*{Acknowledgment}
We thank the organizers of RADCOR 2011 for their kind hospitality.


\begin{thebibliography}{99}

 \bibitem{cern}
  Judith Katzy, {\it Recent results from ATLAS};   
  Marco Pieri, {\it Recent results form CMS} (talks given at RADCOR 2011 Symposium, Mamallapuram, India) 
\bibitem{vanderBij:1988fb} 
  J.~J.~van der Bij and E.~W.~N.~Glover,
  Phys.\ Lett.\ B {\bf 206}, 701 (1988).

\bibitem{Glover:1988rg} 
  E.~W.~N.~Glover and J.~J.~van der Bij,
  Nucl.\ Phys.\ B {\bf 321}, 561 (1989).

\bibitem{Glover:1988fe} 
  E.~W.~N.~Glover and J.~J.~van der Bij,
  Phys.\ Lett.\ B {\bf 219}, 488 (1989).

\bibitem{Campbell:2011bn} 
  J.~M.~Campbell, R.~K.~Ellis and C.~Williams,
  JHEP {\bf 1107}, 018 (2011)
  [arXiv:1105.0020 [hep-ph]].

\bibitem{Campbell:2011cu} 
  J.~M.~Campbell, R.~K.~Ellis and C.~Williams,
  JHEP {\bf 1110}, 005 (2011)
  [arXiv:1107.5569 [hep-ph]].

\bibitem{deFlorian:1999tp} 
  D.~de Florian and Z.~Kunszt,
  Phys.\ Lett.\ B {\bf 460}, 184 (1999)
  [hep-ph/9905283].

\bibitem{Agrawal:1998ch} 
  P.~Agrawal and G.~Ladinsky,
  Phys.\ Rev.\ D {\bf 63}, 117504 (2001)
  [hep-ph/0011346].

\bibitem{Vermaseren:2000nd}
  J.~A.~M.~Vermaseren,~
  {\tt math-ph/0010025}.


\bibitem{vanOldenborgh:1989wn}
  G.~J.~van Oldenborgh and J.~A.~M.~Vermaseren,
  \emph{Z.\ Phys.\  C} {\bf 46} (1990) 425.

\bibitem{vanNeerven:1983vr} 
  W.~L.~van Neerven and J.~A.~M.~Vermaseren,
  Phys.\ Lett.\ B {\bf 137}, 241 (1984).


\bibitem{vanHameren:2010cp} 
  A.~van Hameren,
  Comput.\ Phys.\ Commun.\  {\bf 182}, 2427 (2011)
  [arXiv:1007.4716 [hep-ph]].


\bibitem{'tHooft:1978xw}
  G.~'t Hooft and M.~J.~G.~Veltman,
  Nucl.\ Phys.\  B {\bf 153}, 365 (1979).

\bibitem{Duplancic:2000sk} 
  G.~Duplancic and B.~Nizic,
  Eur.\ Phys.\ J.\ C {\bf 20}, 357 (2001)
  [hep-ph/0006249].

\bibitem{Shivaji:2010aq} 
  A.~Shivaji,
  arXiv:1008.4792 [hep-ph].

\bibitem{Appelquist:1974tg}
  T.~Appelquist and J.~Carazzone,
  Phys.\ Rev.\  D {\bf 11}, 2856 (1975).

\bibitem{Veseli:1997hr} 
  S.~Veseli,
  Comput.\ Phys.\ Commun.\  {\bf 108}, 9 (1998).

\bibitem{pvm}
Message Passing Interface Forum, Int. J. Supercomp. Apps. 8, {\bf 157} (1994)

\end{thebibliography}
\end{document}